\documentclass[reprint,superscriptaddress,preprintnumbers,amsmath,amssymb,aps,prb,floatfix]{revtex4-2}

\usepackage[dvips]{graphicx}
\usepackage{dcolumn}
\usepackage{bm}
\usepackage[svgnames,psnames]{xcolor}
\usepackage[colorlinks,citecolor=DarkGreen,linkcolor=FireBrick]{hyperref}

\newcommand{\Kcpl}[0]{K_\mathrm{cpl}}
\usepackage{physics}

\usepackage{orcidlink}
\begin{document}

\title{Binary-coupling sparse SYK: an improved model of quantum chaos and holography}

\preprint{DMUS-MP-22/17, RIKEN-iTHEMS-Report-22}

\author{Masaki Tezuka \orcidlink{0000-0001-7877-0839}}
 \email{tezuka@scphys.kyoto-u.ac.jp}
\affiliation{Department of Physics, Kyoto University, Kitashirakawa, Sakyo-ku, Kyoto 606-8502, Japan}

\author{Onur Oktay \orcidlink{0000-0001-8408-937X}}
\affiliation{Department of Mathematics, University of Surrey, Guildford, Surrey, GU2 7XH, United Kingdom}

\author{Enrico Rinaldi \orcidlink{0000-0003-4134-809X}}
\affiliation{Physics Department, University of Michigan, Ann Arbor, MI 48109, United States}
\affiliation{Theoretical Quantum Physics Laboratory, Cluster of Pioneering Research, RIKEN, Wako, Saitama 351-0198, Japan}
\affiliation{Interdisciplinary Theoretical \& Mathematical Science Program (iTHEMS), RIKEN, Wako, Saitama 351-0198, Japan}
\affiliation{Center for Quantum Computing (RQC), RIKEN, Wako, Saitama 351-0198, Japan}

\author{Masanori Hanada \orcidlink{0000-0001-5174-2571}}
\affiliation{Department of Mathematics, University of Surrey, Guildford, Surrey, GU2 7XH, United Kingdom}

\author{Franco Nori \orcidlink{0000-0003-3682-7432}}
\affiliation{Center for Quantum Computing (RQC), RIKEN, Wako, Saitama 351-0198, Japan}
\affiliation{Theoretical Quantum Physics Laboratory, Cluster of Pioneering Research, RIKEN, Wako, Saitama 351-0198, Japan}
\affiliation{Physics Department, University of Michigan, Ann Arbor, MI 48109, United States}
\date{\today}

\begin{abstract}
The sparse version of the Sachdev--Ye--Kitaev (SYK) model reproduces essential features of the original SYK model while reducing the number of disorder parameters.
In this paper, we propose a further simplification of the model which we call the binary-coupling sparse SYK model.
We set the nonzero couplings to be $\pm 1$, rather than being sampled from a continuous distribution such as Gaussian. 
Remarkably, this simplification turns out to be an improvement: the binary-coupling model exhibits strong correlations in the spectrum, which is the important feature of the original SYK model that leads to the quick onset of the random-matrix universality, more efficiently in terms of the number of nonzero terms.
This model is better suited for analog or digital quantum simulations of quantum chaotic behavior and holographic metals due to its simplicity and scaling properties.
\end{abstract}

\maketitle

\section{Introduction}
The Sachdev--Ye--Kitaev (SYK) model~\cite{Sachdev-Ye:PRL1993, Kitaev1,Kitaev2,Sachdev:PhysRevX.5.041025} in which $N$ fermions have all-to-all $q$-fermion random couplings has been studied intensively in the last six years.
Though the Hamiltonian is very simple, this model exhibits rich physics and serves as an important toy model for quantum chaos and holography~\cite{Maldacena-Stanford:PhysRevD.94.106002,Maldacena:2019ufo,Trunin_2021,Chowdhury-Georges-Parcollet-Sachdev-2021,Bousso:2022ntt,Faulkner:2022mlp,Catterall:2022wjq}.
It is also a good target for quantum simulation in the near future, which can be the first step toward laboratory experiments for quantum gravitational systems via holographic duality~\cite{Danshita:2016xbo,Franz:2018cqi,Jafferis:2022crx}.
Moreover, the SYK model and its variants have been used in condensed matter as a candidate overarching theory to describe \emph{strange metals} (non-Fermi liquids) and their behavior~\cite{Sachdev:2010um,Song:2017pfw}.
Given the importance of the SYK model for understanding strongly-correlated and disordered systems, a natural question is: Can we simplify the model further while keeping the essential properties leading to its complex and rich physical behavior? 
There are at least a few motivations to ask such a question, including: (i) we want to know which feature of SYK is important for chaos, and (ii) simpler models may admit simpler experimental realizations or quantum simulation protocols. 

Simplifications that retain all the main features of the full, and more cumbersome to study, SYK model may come from thinking about the role of the $q$-fermion interactions. 
Firstly, the all-to-all nature of the interaction may not be necessary; perhaps, some of the couplings can be turned off.
Such a sparse version of the SYK model was proposed and discussed in Refs.~\cite{xu2020sparse, garciagarcia2020sparse}.
The spectral form factor of the model has been studied in Ref.~\cite{Caceres-Misobuchi-Raz:2204.07194}.
Alternatively, one may relax the Gaussian condition on the random couplings.
The effect of non-Gaussian random couplings has been studied in Ref.~\cite{Krajewski_2019}.

In this paper, we propose a novel parsimonious version of the SYK model, the \textit{binary-coupling SYK model}.
We set many couplings to be zero as in the sparse SYK model, and set nonzero couplings to be binary, just $+1$ or $-1$.
\footnote{The idea to use binary couplings appeared as early as in \cite{Kitaev1}, though quantitative comparison between Gaussian and binary couplings has not been made to the best of the present authors' knowledge.}
We study the spectral statistics of the energy spectrum, including the gap ratio and the spectral form factor, which are crucial to determine the connection with random-matrix universality.
Our main result is that this model generates correlations in the spectrum more efficiently, in terms of the number of nonzero terms. Specifically, we observe a stronger spectral rigidity, or equivalently, a quicker onset of the random-matrix universality in the spectral form factor.

We present our results for the Majorana-fermion version of the SYK model with four-fermion couplings.
Other variants of the model have also been extensively studied in the literature, such as the complex fermion~\cite{Sachdev:PhysRevX.5.041025,Louw-Kehrein-PhysRevB.105.075117,Zanoci-Swingle-PhysRevB.105.235131,Davison:2016ngz,Gu:2019jub}
(also known as the two-body random ensemble since 1970s~\cite{French1970,*Bohigas1971}),
bosonic~\cite{PhysRevResearch.2.033025,Fu:2016yrv},
large $S$ spin~\cite{PhysRevB.100.155128},
multiflavor fermions~\cite{Gross_2017},
supersymmetric~\cite{Fu:2016vas,Li_2017,PhysRevLett.124.244101,Gates_2021a,*Gates_2021b},
and nonhermitian versions~\cite{Garcia-Garcia:2021rle},
as well as the two-site~\cite{Maldacena:2018lmt,Jia-Rosa-Verbaarschot2022}
and lattice versions of the model~\cite{Berkooz:2016cvq,Gu_2017},
and SYK-like models with
Yukawa coupling~\cite{PhysRevB.103.195108},
and electron-phonon coupling~\cite{Guo:2019csw,Esterlis_2019}. 
Our approach may generalize to a large portion of such systems.
Because of the simplicity and good scaling with $N$, the use of the binary-coupling models should make simulations more tractable. 

This paper is organized as follows.
In Sec.~\ref{sec:Model}, we define the model, discuss the choice of the sign of the coupling $\pm 1$, and explain that the number of possible realizations of the model is finite but very large.
In Sec.~\ref{sec:Results}, we present our numerical results based on exact diagonalization.
Then we summarize the paper.

\section{Model}\label{sec:Model}
The Hamiltonian of the Majorana-type SYK model with $q$-fermion interactions is written as
\begin{align}
    \mathcal{H}_{\mathrm{SYK}_q} = i^\frac{q}{2}\sum_{1\leq a_1 < \cdots < a_q\leq N}J_{a_1a_2\ldots a_q}\chi_{a_1}\chi_{a_2}\cdots\chi_{a_q},
\end{align}
in which a set of $N$ Majorana fermions $\chi_{1,2,\ldots,N}$ have the anticommutation relation
\begin{equation}
\{\chi_a,\chi_b\}\equiv \chi_a\chi_b + \chi_b\chi_a = 2\delta_{ab}, \label{eqn:MajoranaNormalization}
\end{equation}
and the $N_\mathrm{total}\equiv \begin{pmatrix}N\\q\end{pmatrix}$ couplings $\qty{J_{a_1a_2\ldots a_q}}$ obey the Gaussian distribution~\cite{Kitaev1,Kitaev2,Sachdev-Ye:PRL1993}.
We assume $q$ is even and $J_{a_1a_2\ldots a_q}$ is real. 

The sparse SYK model \cite{xu2020sparse, garciagarcia2020sparse} is defined by setting $J_{a_1a_2\ldots a_q}$ to be zero at a probability $(1-p)$.
Then the number of couplings left nonzero is
\begin{align}
    \Kcpl\equiv p N_\mathrm{total}.
\end{align}
Let us call this model the \textit{Gaussian sparse SYK model}, to distinguish it from the model we propose below. 
Typically, the number of nonzero couplings is chosen to be $\Kcpl=\mathcal{O}(N)$.
Let us also call the original model that corresponds to $p=1$ as the \textit{Gaussian dense SYK model}. 

In the following, we present our proposal for the $q=4$ case for brevity. Generalization to $q\geq6$ is straightforward.
Here we study the Hamiltonian
\begin{align}
\mathcal{H} = C_{N,p} \sum_{(a,b,c,d)}J_{abcd}\chi_a\chi_b\chi_c\chi_d,
\label{eqn:SYK}
\end{align}
in which the summation is understood to be over choices of $(a,b,c,d)$ such that $1\leq a < b < c < d\leq N$, and
$J_{abcd}$ is $+1$ with probability $p/2$, $-1$ with probability $p/2$, and $0$ with probability $(1-p)$.
The normalization constant $C_{N,p}$ will be explained shortly. 
We call this model the \textit{binary-coupling sparse SYK model} because nonzero couplings take only two values. 
We can simplify the model further and define the \textit{unary-coupling sparse SYK model} by taking $J_{abcd}$ as $+1$ with probability $p$ and as $0$ with probability $(1-p)$.
We will discuss the unary-coupling model in Section~\ref{sec:all+1} of the Supplemental Materials.

In the main text, we will focus on the binary-coupling model.
We assume $\Kcpl$ is even.
Rather than generating the value of each $J_{abcd}$ independently of others for an ensemble of models having varying numbers of $+1$ and $-1$, we randomly choose $\Kcpl/2$ couplings to be set to $+1$ and randomly choose $\Kcpl/2$ couplings from the remainder to $-1$.
\subsection{Normalization constant \texorpdfstring{$C_{N,p}$}{C(N,p)}}
The overall scaling constant $C_{N,p}$ does not affect the spectral statistics. 
In the following, $C_{N,p}$ is chosen so that the variance of the energy eigenvalues equals unity. 
The sum of the square of the energy eigenvalues $\qty{\epsilon_j}$ of the $2^{N/2}$-dimensional Hamiltonian $\mathcal{H}$ is obtained as
\begin{align}
    \sum_j \epsilon_j^2=\Tr \mathcal{H}^2=C_{N,p}^2 2^{N/2}\sum_{(a,b,c,d)}J_{abcd}^2 .
\label{eqn:en-eig-sq}
\end{align}

Thus, when $J_{abcd} = \pm 1$ for $pN_\mathrm{total}$ choices of $(a,b,c,d)$, the variance of the $\qty{\epsilon_j}$ equals unity for $C_{N,p}=1/\sqrt{\Kcpl}$.
Note that in many other publications for dense SYK, including \cite{Cotler_2017}, the variance of $J_{abcd}$ scales as $N^{-3}$, and the normalization of the Majorana fermions is often half of \eqref{eqn:MajoranaNormalization}, so that the variance of $\qty{\epsilon_j}$ takes a different form, $(N-1)(N-2)(N-3)/(64N^2)\sim N$.
\subsection{The finiteness of the possible realizations}
There are infinitely many realizations of the Gaussian dense and sparse SYK models because there are infinitely many choices of disorder parameters. 
On the other hand, there are only a finite number of realizations of the binary-coupling model at each fixed $N$. 
Still, the rapid growth of $N_\mathrm{total}\sim N^4/4!$ as a function of $N$ allows us to generate a very large number of distinct models even for $N\sim 10$. 
Practically, the finite size of the realization ensemble is not an issue. 
\section{Spectral statistics}\label{sec:Results}
The Bohigas--Giannoni--Schmit conjecture~\cite{BGS_PhysRevLett.52.1} and later studies stated that, for quantum mechanical Hamiltonians corresponding to classical systems with chaotic behavior,
the spectral statistics of the energy eigenvalues obey that of a Random Matrix Theory (RMT) having the same symmetry of the Hamiltonian.
For Hamiltonians without any symmetry, the random matrix ensemble is the Gaussian unitary ensemble (GUE),
whereas when the anti-unitary time-reversal operator $T$ commutes with the Hamiltonian, the ensemble is the Gaussian orthogonal (resp., symplectic) ensemble, GOE (resp., GSE), if $T^2=1$ (resp., $-1$).
This has been extended to quantum systems without direct correspondence to classical systems.
The Gaussian dense and sparse SYK models exhibit excellent agreement with the RMT~\cite{Cotler_2017,Gharibyan_2018, garciagarcia2020sparse,Caceres-Misobuchi-Raz:2204.07194}, namely GOE for $N\equiv 0~\mathrm{mod}~8$, GUE for $N\equiv 2,6~\mathrm{mod}~8$, and GSE for $N\equiv 4~\mathrm{mod}~8$.
\begin{figure}
    \centering
    \includegraphics{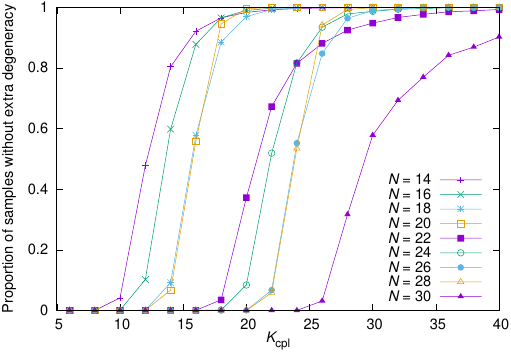}
    \caption{The observed ratio of samples with least degeneracy within a parity sector (no degeneracy for $N\not\equiv4~\mathrm{mod}~8$, twofold degeneracy for $N\equiv4~\mathrm{mod}~8$), plotted against the number of nonzero couplings $\Kcpl$.
    Half of the couplings are positive and the rest are negative.
    The number of samples is chosen so that $2^{24}$ eigenvalues are used for analysis for each $(N, \Kcpl)$.
    }
    \label{fig:pm_Nm-degen-parity-single-plot}
\end{figure}
\begin{figure}
    \centering
    \includegraphics{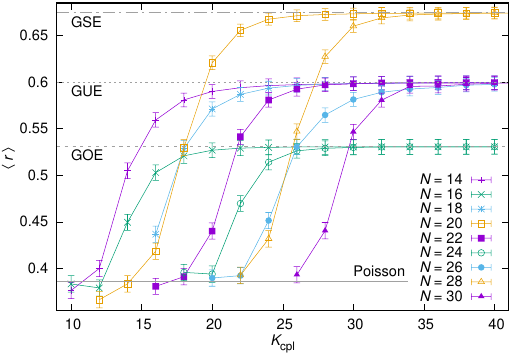}
    \caption{The average of neighboring gap ratio between non-degenerate eigenvalues for samples with least degeneracy in Fig.~\ref{fig:pm_Nm-degen-parity-single-plot}, plotted against the number of nonzero couplings. ``GSE'', ``GUE'', ``GOE'', and ``Poisson'' indicate the values in Ref.~\cite{Atas2013-PhysRevLett.110.084101}.
    }
    \label{fig:pm_Nm-r-avg}
\end{figure}

\subsection{Extra degeneracy for smaller numbers of couplings}
In the Gaussian dense SYK model with $q=4$, the eigenvalues are 
\begin{itemize}
    \item not degenerate in general for $N\equiv 0~\mathrm{mod}~8$.
    \item doubly degenerate due to the bijective map between the two parity sectors for $N\equiv 2, 6~\mathrm{mod}~8$.
    \item doubly degenerate within each parity sector, but not degenerate in general between the parity sectors for $N\equiv 4~\mathrm{mod}~8$.
\end{itemize}

With only a very small number of couplings, there may be extra degeneracy due to some accidental symmetries, as have been discussed for the Gaussian sparse SYK model~\cite{garciagarcia2020sparse}. Such accidental symmetries increase the degeneracy of some of the energy eigenvalues. Therefore, for small values of $\Kcpl$, high orders of degeneracy is observed. In some samples, all eigenvalues have the same higher degeneracy such as four. In other samples, the eigenvalues have varying orders of degeneracy within one sample. For $\Kcpl\gtrsim N$, the degeneracy is uniform within each sample and the order is a power of two. Still, extra degeneracy is observed for some samples. At sufficiently large $\Kcpl$, samples with extra degeneracy becomes rarer. 
The probability of observing samples without extra degeneracy is plotted in Fig.~\ref{fig:pm_Nm-degen-parity-single-plot}.
While how the probability approaches unity depends on $N~\mathrm{mod}~8$, we observe that for $N=14, 16, \ldots, 30$, more than half of the samples are without extra degeneracy if the number of couplings $\Kcpl$ exceeds $N$.
\subsection{Comparison with Random Matrix Theory}
A standard criterion to establish quantum chaos is to confirm the random-matrix behavior of the energy spectrum.
Not surprisingly, the binary-coupling SYK model passes this test. 
In this subsection, we consider the average, over many realizations, of the binary-coupling SYK model. If $N$ is sufficiently large, a good agreement with RMT can be observed for each realization; see Sec.~\ref{sec:singleRealizations} of the Supplemental Materials regarding this point.

In Fig.~\ref{fig:pm_Nm-r-avg}, the average of neighboring nonzero gap ratio $r_i\equiv{\mathrm{min}(s_i,s_{i+1})}/{\mathrm{max}(s_i,s_{i+1})}$, where $s_i\equiv E_{i+1}-E_i$ for the ordered set of distinct eigenvalues $\{E_i\}\subseteq \{\epsilon_j\}$, is plotted for samples with least degeneracy within each parity sector.
As we have seen in the above, for number of couplings $\gtrsim N$, most samples do not have extra degeneracy.
For such cases, the value of $\langle r\rangle$ rapidly approaches that of RMT.

By using the partition function analytically continued as a function of the inverse temperature $\beta$ to $\beta + it$ with $t\in \mathbb{R}$,
\begin{align}
    Z(t,\beta)\equiv Z(\beta+it) = \sum_j \exp\left(-(\beta+it)\epsilon_j\right),
\end{align}
we define the spectral form factor $g(t,\beta)$ as
\begin{align}
    g(t,\beta)=\vert Z(t,\beta)\vert^2/Z(0,\beta)^2. 
\end{align}
The late-time features of the spectral form factor, specifically the ramp and plateau, are useful criteria to see the universality described by RMT.

\begin{figure}
    \centering
    \includegraphics{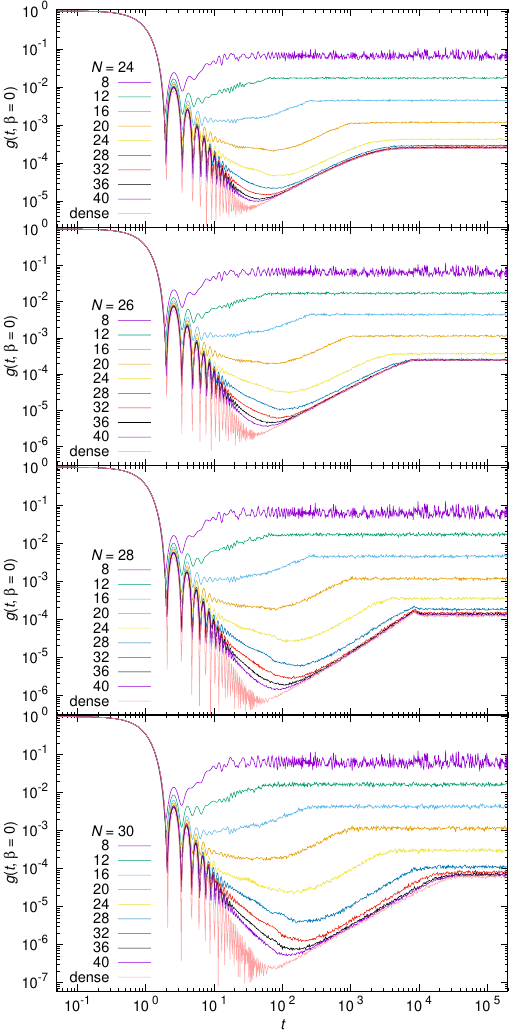}
    \caption{
    $g(t,\beta)=\vert Z(t,\beta)\vert^2/Z(0,\beta)^2$ for $Z(t,\beta)=Z(\beta+it)=\sum_j \exp(-(\beta+it)\epsilon_j)$ versus time $t$ for the binary sparse SYK model. $\beta=0$.
    The value of $N$ as well as the number of nonzero couplings, $\Kcpl$, are indicated in the legend for each plot.
    The shift of the height of the plateau for $\Kcpl\lesssim N$ is due to the extra degeneracy of the energy eigenvalues.
    }
    \label{fig:N24-30-Gt}
\end{figure}

In Fig.~\ref{fig:N24-30-Gt}, we plotted the spectral form factor for $N=24,26,28,30$. Late-time behaviors agree with those of GOE ($N=24$), GUE ($N=26,30$), GSE ($N=28$) random matrices: we observe a long ramp proportional to $t^{1}$ as $\Kcpl$ is increased.

\subsection{Comparison with Gaussian sparse SYK}
So far, we have seen that the binary-coupling sparse SYK model exhibits quantum chaos. Now we show that the binary model generates correlation in the spectrum more efficiently in terms of the number of nonzero terms. Specifically, we observe the stronger spectral rigidity for the same values of $N$ and $p$. This means the binary model is more chaotic than the Gaussian model, in the sense that the energy spectrum is closer to RMT. 

Let us consider the unfolded energy spectrum. (Regarding the unfolding, see Ref.~\cite{Bohigas-1989,*Guhr_1998,*Haake-2001,*Gomez-PhysRevE.66.036209}.)
Let $\bar{\Delta}$ be the average nearest-neighbor level spacing. For $K>0$, we denote the number of energy eigenvalues between $E$ and $E+K\bar{\Delta}$ by $n(E,K)$. The number variance $\Sigma^2(K)$ is defined by
$\Sigma^2(K)=\langle n^2(E,K)\rangle-\langle n(E,K)\rangle^2$, 
where $\langle\ \cdot\ \rangle$ denote the average over $E$. We can compare the number variance with RMT. If the agreement is observed up to a larger value of $K$, then the system is more strongly chaotic in the sense that the Random-Matrix universality sets in at earlier time scales. 
RMT shows rather small $\Sigma^2(K)$, which means the spectrum is rigid due to the level repulsion. This property is called the spectral rigidity.

\begin{figure}
    \centering
    \includegraphics{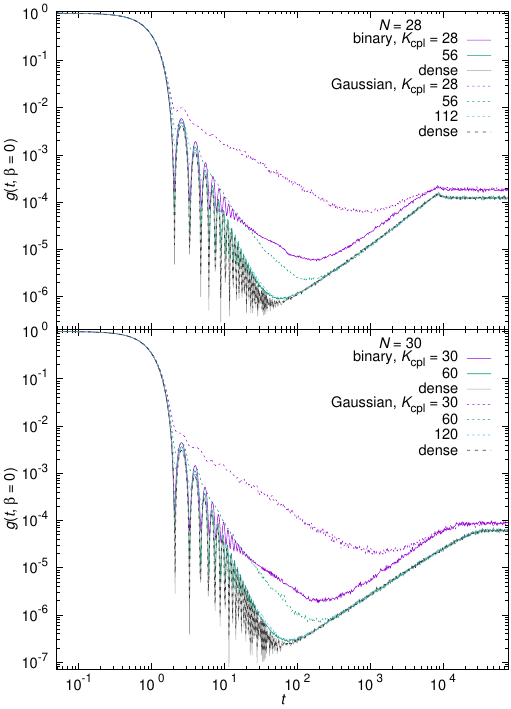}
    \caption{
    Spectral form factors $g(t,\beta=0)$ versus time $t$, for the binary-coupling SYK and Gaussian SYK, for $N=28$ and $N=30$. }
    \label{fig:SFF_binary_gaussian_g}
\end{figure}

While the number variance provides us with a simple way to detect the spectral rigidity, it may depend on the details of the unfolding procedures~\cite{Gharibyan_2018}. The spectral form factor gives another way to see the spectral rigidity which is not sensitive to the details of the unfolding; if the spectral rigidity is stronger, the ramp is longer. 
In Fig.~\ref{fig:SFF_binary_gaussian_g}, we show the spectral form factors for the binary and Gaussian, sparse and dense SYK models with $N=28$ and $N=30$.
The ramp starts significantly earlier for the binary model with $N$ and $2N$ couplings compared to the Gaussian model with the same number of couplings.
The ramp for the binary model with $2N$ couplings is as long as that of the Gaussian model with $4N$ couplings.
Finally, in the dense SYK limit, binary and Gaussian models are indistinguishable.

\begin{figure}
    \centering
    \includegraphics{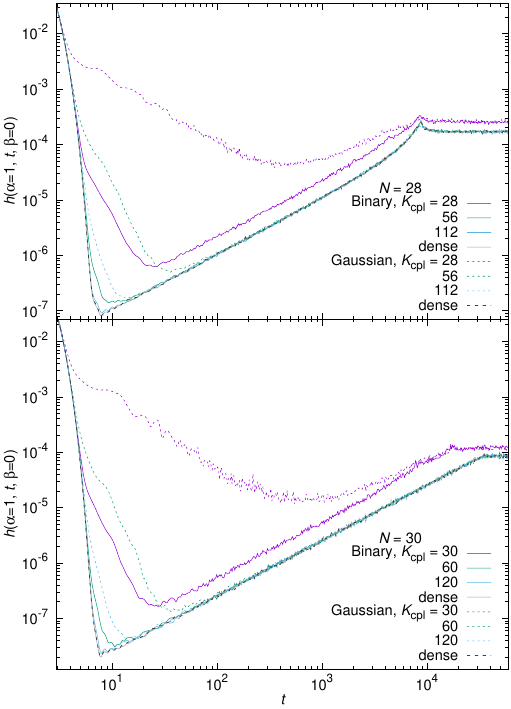}
    \caption{
    The modified version of the spectral form factor $h(\alpha=1,t,\beta=0)$ versus time $t$, for the binary-coupling SYK and Gaussian SYK, for $N=28$ and $N=30$. }
    \label{fig:SFF_binary_gaussian_h}
\end{figure}

Strictly speaking, the onset of the ramp can be hidden if the decay of the spectral form factor at early time is not sufficiently fast. Note that the early-time decay (also known as slope) does not reflect the microscopic properties of the energy spectrum. To see the onset of the ramp more accurately, a modified spectral form factor $h(\alpha,t,\beta)$ defined by
\begin{align}
    h(\alpha,t,\beta)=\vert Y(\alpha,t,\beta)\vert^2/Y(0,\beta)^2,  
\end{align}
where
\begin{align}
    Y(\alpha,t,\beta)\equiv \sum_j \exp\left(-\alpha\epsilon_j^2-(\beta+it)\epsilon_j\right),
\end{align}
is useful~\cite{DouglasStanford_2018,Gharibyan_2018}. By choosing an appropriate value of $\alpha$, the slope can be largely removed and the ramp can be seen from the earlier time. In Fig.~\ref{fig:SFF_binary_gaussian_h}, we plot $h(\alpha,t,\beta)$ for $\alpha=1$ and $\beta=0$. Again, the longer ramp can be seen for the binary-coupling model than the Gaussian model, for the same number of the couplings. 

\section{Conclusion and discussion}\label{sec:Summary}
In this paper, we proposed the binary-coupling sparse SYK model. We demonstrated that the spectral statistics obey the RMT predictions for $\mathcal{O}(N)$ number of couplings as in the Gaussian sparse SYK models \cite{xu2020sparse,garciagarcia2020sparse,Caceres-Misobuchi-Raz:2204.07194}. With the same number of couplings, the binary-coupling model shows better agreement with RMT. Therefore, this model is an improvement, rather than just a simplification, of the Gaussian sparse SYK model. 
The better agreement to dense models in the binary case compared to the Gaussian case may be attributed to the high probability of having relatively small coupling amplitudes in the latter.
If we approximate small couplings by zero, effectively the value of $p$ goes down in the Gaussian case.

We envisage the generalization of the present approach to other variants of the SYK model in a similar manner. For example, we may choose the coupling constants from the unit circle on the complex plane, or further limit the values of the couplings to $\{e^{2\pi j/Q}\}_{j=0,1,\ldots,Q-1}$, in which $Q\geq 3$ is an integer, to obtain a version of (sparse or dense) non-hermitian SYK model.

The binary nature of the couplings may simplify the physical realization of the model via digital or analog quantum simulations~\cite{Buluta-Nori-2009,Georgescu-Ashhab-Nori-2014,Luo:2017bno,PRXQuantum.2.017003}. 
It would be nice if experiments on SYK models with sufficiently large $N$ are achieved and ``quantum gravity in the lab''~\cite{Danshita:2016xbo,Franz:2018cqi,https://doi.org/10.48550/arxiv.1911.06314,Shapoval:2022xeo,Jafferis:2022crx} becomes reality.
Simulations on optical Kagome lattices~\cite{Wei:2020ryt} may also be simplified in the binary-coupling sparse SYK, and could lead to tractable studies of high-temperature cuprate superconductors~\cite{Sachdev:2010um}.

\vspace*{1cm}
\begin{acknowledgments}
The authors thank Brian Swingle for helpful comments.
M.~T. thanks Antonio M. Garc\'ia-Garc\'ia, Chisa Hotta, and Norihiro Iizuka for discussions; Yoshifumi Nakata and Satyam Shekhar Jha for collaborations on related projects.

M.T. was partially supported by the Japan Society for the Promotion of Science (JSPS) Grants-in-Aid for Scientific Research (KAKENHI) Grants No. JP17K17822, JP20K03787, JP20H05270, and JP21H05185. 
O.O. was supported by TUBITAK Grant No. $2219$.
M.H. was supported by the STFC Ernest Rutherford Grant ST/R003599/1. 
E.R. was supported by Nippon Telegraph and Telephone Corporation (NTT) Research.
F.N. is supported in part by: NTT Research, the Japan Science and Technology Agency (JST) [via the Quantum Leap Flagship Program (Q-LEAP), Moonshot R\&D Grant No.~JPMJMS2061], JSPS [via KAKENHI Grant No.~JP20H00134], the Army Research Office (ARO) (Grant No.~W911NF-18-1-0358), the Asian Office of Aerospace Research and Development (AOARD) (via Grant No.~FA2386-20-1-4069), and the Foundational Questions Institute Fund (FQXi) via Grant No.~FQXi-IAF19-06.
The authors thank the Royal Society International Exchanges award IEC/R3/213026.
Part of the numerical computations were performed using the facilities of the Supercomputer Center, the Institute for Solid State Physics, the University of Tokyo.
\end{acknowledgments}
\begin{center}
\textbf{DATA MANAGEMENT}
\end{center}
No additional research data beyond the data presented and cited in this work are needed
to validate the research findings in this work. Simulation data will be publicly available
after publication.

\bibliography{references}

\newpage\hbox{}\thispagestyle{empty}\newpage
\onecolumngrid

\begin{center}
\textbf{\Large Supplemental Materials: Binary-coupling sparse SYK model}
\end{center}
\twocolumngrid
\setcounter{section}{0}
\setcounter{equation}{0}
\setcounter{figure}{0}
\setcounter{table}{0}
\setcounter{page}{1}
\makeatletter
\renewcommand{\thesection}{S\arabic{section}}
\renewcommand{\theequation}{S\arabic{equation}}
\renewcommand{\thefigure}{S\arabic{figure}}
\renewcommand{\bibnumfmt}[1]{[S#1]}
\makeatother

\section{Unary-coupling sparse SYK model}\label{sec:all+1}
As we have stated in Sec.~\ref{sec:Model} of the main text, the unary-coupling sparse SYK model (i.e., all nonzero couplings are $+1$) behaves similarly to the binary-coupling model as long as $1\ll \Kcpl\ll N_\mathrm{total}$. Below, we will show basic results regarding the unary-coupling model. 

In the binary-coupling sparse SYK model, there were two sources of randomness: (i) which coupling is nonzero, and (ii) whether each nonzero coupling is $+1$ or $-1$. In the unary-coupling model, the only source of randomness is (i). Therefore, if $\Kcpl$ is too small or too large, it is far less effective than the binary-coupling model. Specifically, when $\Kcpl=N_{\rm total}$, there is no randomness at all. 

The anti-commuting relation between Majorana fermions \eqref{eqn:MajoranaNormalization} means that reordering the Majorana fermions 
$\qty(\chi_1,\chi_2,\ldots,\chi_N)\mapsto\qty(\chi_{\sigma(1)},\chi_{\sigma(2)},\ldots,\chi_{\sigma(N)})$, where $\sigma$ is a non-unit element of the symmetry group $S_N$,
can flip some of the signs of the interaction.
For $1\ll \Kcpl \ll N_\mathrm{total}$, for a typical choice of the nonzero terms, it would be possible to perform the reordering to the binary-coupling model and make most of the signs to be positive, then we do not expect a big difference between ensembles of the binary-coupling and unary-coupling model realizations.  

\begin{figure}
    \centering
    \includegraphics{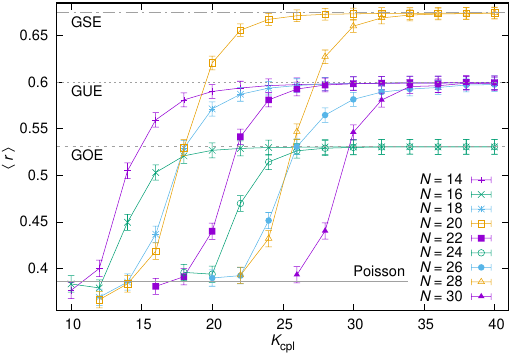}
    \caption{The average of neighboring gap ratio between non-degenerate eigenvalues for unary sparse samples with least degeneracy, plotted against the number of nonzero couplings. ``GSE'', ``GUE'', ``GOE'', and ``Poisson'' indicate the values in Ref.~\cite{Atas2013-PhysRevLett.110.084101}.
    }
    \label{fig:pm_Nm-r-avg-unary}
\end{figure}

\begin{figure*}
    \centering
    \includegraphics{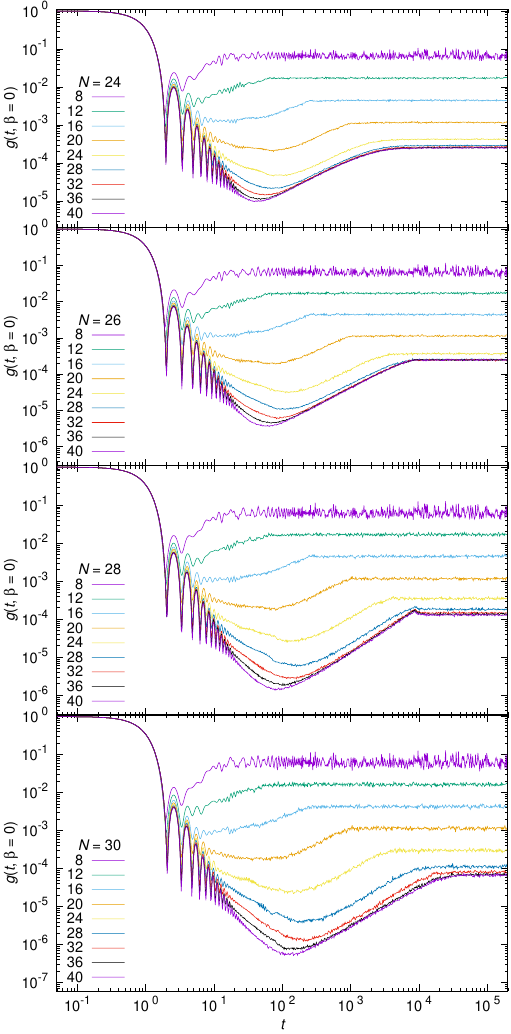}
    \caption{
    The spectral form factor
    $g(t,\beta=0)$ versus time $t$ for the unary-coupling sparse SYK model. The value of $N$ as well as the number of nonzero couplings, $\Kcpl$, are indicated in the legend for each plot.
    }
    \label{fig:N24-30-Gt-unary}
\end{figure*}

In Fig.~\ref{fig:pm_Nm-r-avg-unary}, we plot the average of neighboring gap ratio between non-degenerate eigenvalues for the \textit{unary}-coupling sparse SYK model. The observed values are nearly identical to those for the \textit{binary}-coupling model in Fig.~\ref{fig:pm_Nm-r-avg} in the main text.
In Fig.~\ref{fig:N24-30-Gt-unary}, we plot the spectral form factor for the unary-coupling model. Again, the results exhibit little difference from the binary-coupling model in Fig.~\ref{fig:N24-30-Gt} in the main text.
Note that changing all nonzero $J_{abcd}$ for a particular realization of the binary-coupling model results in a significant change of its spectrum. The observed agreement between the two models is between their ensembles after the averaging.

\section{Example of single realizations for \texorpdfstring{$N=32,34$}{N=32,34}}\label{sec:singleRealizations}
\begin{widetext}
\begin{figure*}
    \centering
    \includegraphics{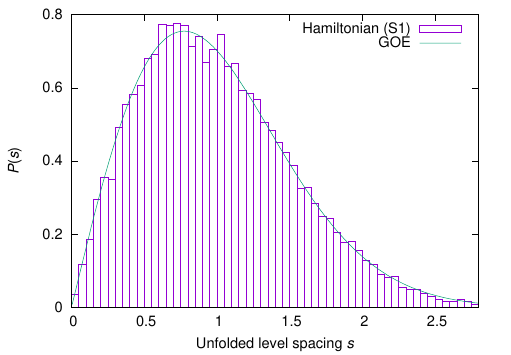}
    \includegraphics{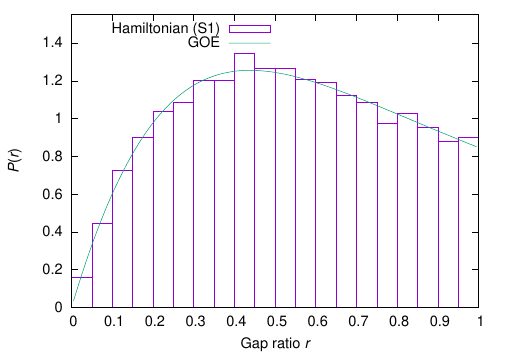}
    \includegraphics{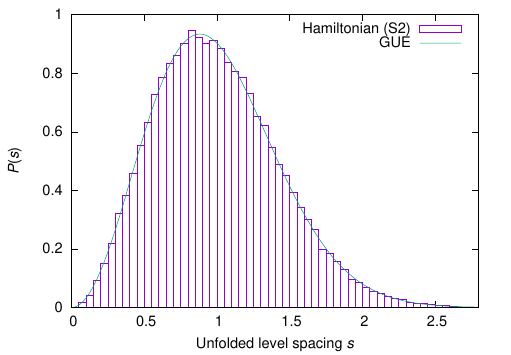}
    \includegraphics{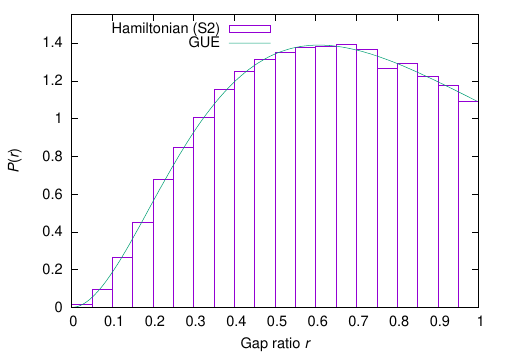}
    \caption{
    The distribution of the nearest-neighbor level spacing $P(s)$ and that of the neighboring gap ratio $P(r)$ for the eigenvalues of the single realization of the binary sparse SYK model. 
    [Top] $N=32$ and $\Kcpl=30$ with the specific realization given by eq.~\eqref{eqn:N32},
    [Bottom] $N=34$ and $\Kcpl=36$ with the specific realization given by eq.~\eqref{eqn:N34}.
    For $P(s)$, we omit the largest $5\%$ and smallest $5\%$ of the eigenvalues from the analysis to prevent the eigenvalues near the edges from affecting the polynomial fit of the spectrum. We use tenth-order polynomial fitting for unfolding the spectrum.
    }
    \label{fig:N32-34}
\end{figure*}
\end{widetext}

In Fig.~\ref{fig:N32-34}, we show the distributions for the unfolded nearest-neighbor level separation $P(s)$ and neighboring gap ratio $P(r)$ for single realizations of the binary-coupling sparse SYK model for $(N,\Kcpl)=(32,30), (34,36)$.
The Hamiltonian we used are 
\begin{align}
    \mathcal{H}&=
    \chi_{1}\chi_{2}\chi_{3}\chi_{4}
    -\chi_{1}\chi_{6}\chi_{10}\chi_{21}
    -\chi_{1}\chi_{8}\chi_{23}\chi_{24}\nonumber\\&
    -\chi_{1}\chi_{11}\chi_{27}\chi_{28}
    +\chi_{1}\chi_{22}\chi_{26}\chi_{27}
    +\chi_{2}\chi_{5}\chi_{10}\chi_{23}\nonumber\\&
    +\chi_{2}\chi_{15}\chi_{25}\chi_{30}
    +\chi_{3}\chi_{5}\chi_{10}\chi_{32}
    -\chi_{3}\chi_{5}\chi_{24}\chi_{31}\nonumber\\&
    +\chi_{3}\chi_{20}\chi_{24}\chi_{26}
    +\chi_{4}\chi_{8}\chi_{18}\chi_{23}
    -\chi_{5}\chi_{10}\chi_{23}\chi_{30}\nonumber\\&
    +\chi_{5}\chi_{19}\chi_{23}\chi_{30}
    -\chi_{5}\chi_{25}\chi_{29}\chi_{32}
    -\chi_{6}\chi_{7}\chi_{20}\chi_{23}\nonumber\\&
    +\chi_{7}\chi_{9}\chi_{12}\chi_{15}
    +\chi_{7}\chi_{10}\chi_{12}\chi_{18}
    -\chi_{7}\chi_{21}\chi_{23}\chi_{27}\nonumber\\&
    -\chi_{7}\chi_{24}\chi_{28}\chi_{31}
    +\chi_{8}\chi_{9}\chi_{15}\chi_{32}
    -\chi_{9}\chi_{15}\chi_{25}\chi_{30}\nonumber\\&
    +\chi_{9}\chi_{19}\chi_{21}\chi_{27}
    +\chi_{10}\chi_{11}\chi_{19}\chi_{32}
    +\chi_{10}\chi_{12}\chi_{14}\chi_{16}\nonumber\\&
    -\chi_{11}\chi_{17}\chi_{25}\chi_{28}
    -\chi_{12}\chi_{14}\chi_{20}\chi_{24}
    -\chi_{12}\chi_{19}\chi_{31}\chi_{32}\nonumber\\&
    +\chi_{12}\chi_{23}\chi_{24}\chi_{30}
    -\chi_{13}\chi_{17}\chi_{21}\chi_{27}
    -\chi_{22}\chi_{23}\chi_{26}\chi_{31},\label{eqn:N32}
\end{align}
for $N=32$ and 
\begin{align}
\mathcal{H}&=
\chi_{1}\chi_{6}\chi_{20}\chi_{28}
+\chi_{1}\chi_{7}\chi_{22}\chi_{24}
-\chi_{1}\chi_{10}\chi_{15}\chi_{25}\nonumber\\&
-\chi_{1}\chi_{15}\chi_{19}\chi_{31}
-\chi_{1}\chi_{15}\chi_{21}\chi_{26}
-\chi_{2}\chi_{3}\chi_{17}\chi_{23}\nonumber\\&
+\chi_{2}\chi_{19}\chi_{23}\chi_{24}
+\chi_{3}\chi_{5}\chi_{6}\chi_{16}
+\chi_{3}\chi_{14}\chi_{17}\chi_{22}\nonumber\\&
+\chi_{3}\chi_{15}\chi_{20}\chi_{25}
+\chi_{3}\chi_{21}\chi_{28}\chi_{34}
+\chi_{3}\chi_{23}\chi_{32}\chi_{33}\nonumber\\&
+\chi_{4}\chi_{5}\chi_{6}\chi_{30}
-\chi_{4}\chi_{9}\chi_{15}\chi_{29}
-\chi_{4}\chi_{9}\chi_{30}\chi_{32}\nonumber\\&
+\chi_{4}\chi_{22}\chi_{27}\chi_{30}
-\chi_{4}\chi_{23}\chi_{26}\chi_{34}
+\chi_{5}\chi_{8}\chi_{14}\chi_{31}\nonumber\\&
-\chi_{5}\chi_{10}\chi_{15}\chi_{18}
-\chi_{6}\chi_{7}\chi_{18}\chi_{30}
+\chi_{6}\chi_{13}\chi_{30}\chi_{32}\nonumber\\&
-\chi_{6}\chi_{14}\chi_{20}\chi_{25}
-\chi_{6}\chi_{15}\chi_{23}\chi_{32}
-\chi_{6}\chi_{18}\chi_{32}\chi_{34}\nonumber\\&
+\chi_{6}\chi_{21}\chi_{31}\chi_{32}
-\chi_{7}\chi_{24}\chi_{28}\chi_{30}
+\chi_{8}\chi_{13}\chi_{14}\chi_{19}\nonumber\\&
+\chi_{9}\chi_{11}\chi_{25}\chi_{29}
-\chi_{10}\chi_{13}\chi_{21}\chi_{34}
+\chi_{11}\chi_{12}\chi_{29}\chi_{33}\nonumber\\&
+\chi_{11}\chi_{22}\chi_{28}\chi_{30}
-\chi_{13}\chi_{21}\chi_{23}\chi_{25}
+\chi_{15}\chi_{18}\chi_{27}\chi_{28}\nonumber\\&
-\chi_{16}\chi_{25}\chi_{27}\chi_{28}
-\chi_{17}\chi_{19}\chi_{24}\chi_{28}
-\chi_{19}\chi_{25}\chi_{31}\chi_{33},\label{eqn:N34}
\end{align}    
for $N=34$. 
The results agree well with those for the GOE and GUE random matrices~\cite{Dietz-Haake-1990,Atas2013-PhysRevLett.110.084101}, respectively.

\end{document}